\documentclass[journal=nalefd,layout=twocolumn,email=true]{achemso}

\usepackage[abs]{overpic}
\usepackage{graphicx,graphics}
\usepackage{dcolumn}
\usepackage{amsmath,amssymb,amsfonts}
\usepackage{braket}
\usepackage{latexsym,verbatim}
\usepackage{bm}
\usepackage{color}
\usepackage[dvipsnames]{xcolor}
\usepackage[normalem]{ulem}
\usepackage[framemethod=default]{mdframed}
\usepackage[unicode=true,breaklinks=true,colorlinks,citecolor=blue,linkcolor=blue,urlcolor=blue]{hyperref}
\usepackage[makeroom]{cancel}
\usepackage{fancybox}
\usepackage{siunitx}

\usepackage{lipsum}
\usepackage{mathtools}
\usepackage{cuted}

\usepackage{multicol}

\definecolor{greenI}{rgb}{0, .4, 0}

\usepackage[version=4]{mhchem}
\usepackage{siunitx}

\usepackage{glossaries}
\setacronymstyle{long-short}
\glsdisablehyper
\usepackage[labelfont=bf]{caption}
\newcommand{\NanoMeter}[1]{\SI{#1}{\nano\meter}}
\newcommand{\NanoAmpere}[1]{\SI{#1}{\nano\ampere}}

\newcommand{\Degree}[1]{\SI{#1}{\degree}}

\newacronym{lpge}{LPGE}{linear photogalvanic effect}
\newacronym{cpge}{CPGE}{circular photogalvanic effect}
\newacronym{cpde}{CPDE}{circular photon drag effect}
\newacronym{lpde}{LPDE}{linear photon drag effect}
\newacronym{tblg}{tBLG}{twisted bilayer graphene}
\newacronym{hbn}{hBN}{hexagonal boron nitride}
\newacronym{pge}{PGE}{photogalvanic effect}
\newacronym{pde}{PDE}{photon drag effect}
\newacronym{cnp}{CNP}{charge neutrality point}
\newacronym{thz}{THz}{terahertz}

\author{Maximilian Otteneder}
\altaffiliation{Contributed equally to this work}
\affiliation{Terahertz Center, University of Regensburg, 93040 Regensburg, Germany}

\author{Stefan Hubmann}
\altaffiliation{Contributed equally to this work}
\affiliation{Terahertz Center, University of Regensburg, 93040 Regensburg, Germany}

\author{Xiaobo Lu}
\affiliation{ICFO - Institut de Ciencies Fotoniques, The Barcelona Institute of Science and Technology, Castelldefels, Barcelona 08860, Spain}

\author{Dmitry A. Kozlov}
\affiliation{Rzhanov Institute of Semiconductor Physics, 630090 Novosibirsk, Russia}

\author{Leonid E. Golub}
\affiliation{Ioffe Institute, 194021 St. Petersburg, Russia}

\author{Kenji Watanabe}
\affiliation{
National Institute of Material Science, 1-1 Namiki, Tsukuba 305-0044, Japan} 

\author{Takashi Taniguchi}
\affiliation{
National Institute of Material Science, 1-1 Namiki, Tsukuba 305-0044, Japan}

\author{Dmitri K. Efetov}
\affiliation{ICFO - Institut de Ciencies Fotoniques, The Barcelona Institute of Science and Technology, Castelldefels, Barcelona 08860, Spain}

\author{Sergey D. Ganichev}
\affiliation{Terahertz Center, University of Regensburg, 93040 Regensburg, Germany}
\email{sergey.ganichev@ur.de}

\title[THz]{Terahertz photogalvanics in twisted bilayer graphene close to the second magic angle}

\let\oldmaketitle\maketitle
\let\maketitle\relax
\AtBeginDocument{}

\begin{document}

\twocolumn[
\begin{@twocolumnfalse}
	\oldmaketitle
	\begin{abstract}
		We report on the observation of  photogalvanic effects in twisted bilayer graphene (tBLG) with a twist angle of 0.6$^\circ$. We show that excitation of tBLG bulk causes a photocurrent, whose sign and magnitude are controlled by orientation of the radiation electric field and the photon helicity. The observed photocurrent provides evidence for the reduction of the point group symmetry in low twist-angle tBLG to the lowest possible one. The developed theory shows that the current is formed by asymmetric scattering in gyrotropic tBLG. We also detected the photogalvanic current formed in the vicinity of the edges. For both, bulk and edge photocurrents, we demonstrate the emergence of pronounced oscillations upon variation of the gate voltage. The gate voltages associated with the oscillations coincide well with peaks in resistance measurements. These are well explained by inter-band transitions between a multitude of isolated bands in tBLG.
	\end{abstract}
\end{@twocolumnfalse}
]



\textbf{Introduction.} Twisted bilayer graphene has emerged as one of the richest and most tunable systems in condensed matter physics, displaying a multitude of symmetry broken states such as correlated insulators, superconductors, magnets and topological systems. Consisting of two vertically stacked graphene sheets that are slightly twisted with respect to one another by a twist-angle  $\theta$, a large scale moiré super-potential is created between the layers, which strongly modifies its electronic, optical, as well as the mechanical properties, see e.g.~[\!\!\!\citenum{Santos2007,Li2009,Mele2010,Li2010,Morell2010,Bistritzer2011,Kim2017a,Cao2018a,Cao2018,Sunku2018,Yoo2019,Jiang2019,Yankowitz2019,Lu2019,Hesp2019}]. Strikingly, for tBLG with small twist-angles $\theta$ between $0.1^\circ$ and  $ 1^\circ$, which results in an exceptionally large moiré unit cell, the energy spectrum becomes quite complex with a multitude of flat mini-bands. Furthermore, corrugation and strain effects strongly modify the stacking sequence in each unit cell, leading to the formation of an intricate network of triangular AB and BA regions, which are separated by topologically non-trivial domain walls~\cite{Kim2017a}. 

Recent transport studies have already revealed some of these attributes of small angle \gls{tblg}, however, at this point in time, studies of their optical and opto-electronic properties are still not existent. Transport phenomena, which scale with the second power of the high frequency electric field (see e.g. review~[\!\!\!\citenum{Glazov2014}]), open up new opportunities to study electron transport in \gls{tblg}. In particular, the unusual morphology and electronic spectrum of small angle \gls{tblg} opens up entirely new access to the investigation of the photogalvanic (PGE) and photon drag (at low frequencies also called dynamic Hall) effects, which have been previously observed in single and AB bi-layer graphene~\cite{Glazov2014,Karch2010,Karch2011,Jiang2011,Olbrich2013a,Maysonnave2014,Obraztsov2014,Inglot2015,Hipolito2016,Ganichev2017,Zhu2017,Plank2018a,Candussio2020} and already found applications. For \gls{tblg}, however, such studies have not yet been carried out. 

Here we report on the observation of the PGE excited in the bulk and edges of \gls{tblg}. For linearly polarized radiation the photocurrent magnitudes and directions are controlled by the orientation of the electric field vector, whereas for circularly polarized radiation the current direction is defined by the photon helicity and reverses by switching from right- to left-handed circularly polarized radiation and vice versa. The observed effects show that the surface symmetry of small-angle twisted graphene is reduced as compared to the one expected from regular \gls{tblg}. Variation of the back gate voltage reveals that the photocurrent magnitude is an oscillating function of the gate voltage. Transport measurements carried out in parallel to photocurrent measurements demonstrate that the observed oscillations correlate with the oscillation of the sample resistance. We demonstrate that the PGE is caused by asymmetric scattering of electrons driven by the \gls{thz} electric field. The developed phenomenological and microscopic theories describe the observed photocurrents well. In particular, we show that the photocurrent oscillations as a function of the gate voltage are caused by the shifting of the Fermi energy across  well separated almost flat bands, which in turn results in an oscillation of the density of states.

\begin{figure*}
	\centering
	\includegraphics[width=\linewidth]{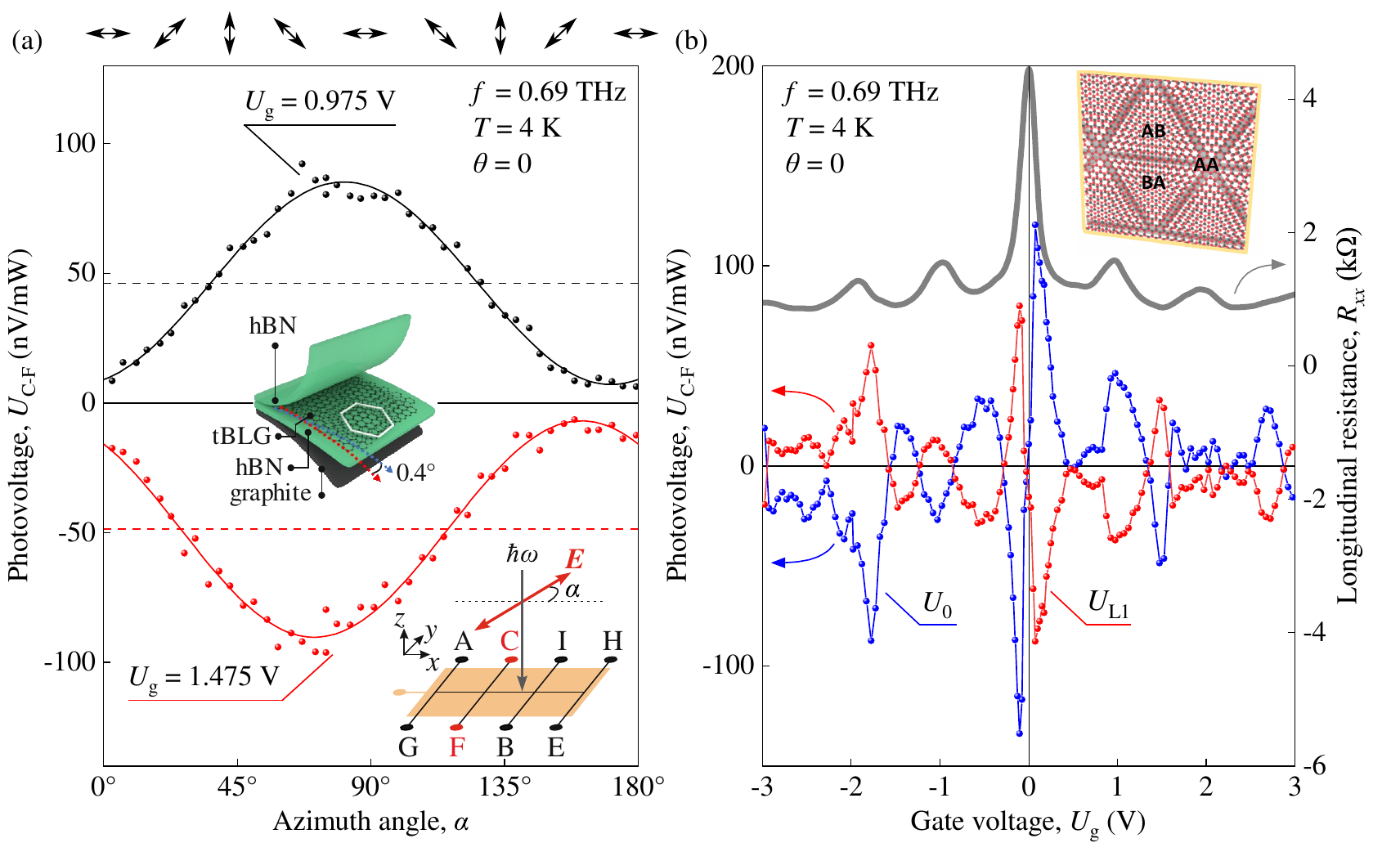}
	\caption{Panel (a): Dependence of the photovoltage on the azimuth angle $\alpha$ measured over contacts C and F  for applied gate voltages of \SI{0.975}{\volt} and \SI{1.475}{\volt}. Sample and contact geometry as well as experimental setup are shown in the bottom right inset. Solid lines are fits after Eqs.~\ref{fit} or theoretical Eqs.~\ref{phenom} and \ref{j_fin} yielding the same polarization dependence. The dashed lines show the offset photovoltage $U_0$. Arrows on top of both panels illustrate the states of polarization for several azimuth angles $\alpha$. The central inset shows a sketch of the sample layers. Panel (b): Red and blue traces (left axis) show the dependence of the parameters $U_{\text L1}$ and $U_0$ on the gate voltage for the bulk contribution, respectively. The gray curve depicts the longitudinal resistance $R_{xx}$. The inset shows a schematic of \gls{tblg} in real space.
	}
	\label{Fig1}
\end{figure*}

\textbf{Devices and measurements.} The hBN/tBLG/hBN/graphite heterostructure was prepared with a ''tear and stack'' van der Waals assembly technique~\cite{Kim2017a}. The two pieces of graphene were rotated by a slightly larger angle of about $ 1^\circ$ and finally relaxed into the twist angle $\approx 0.6^\circ$  during the assembling process. The resulting stack consists of a \NanoMeter{20} \gls{hbn} top layer, \gls{tblg}, \NanoMeter{16} bottom \gls{hbn} and \NanoMeter{3} graphite, which also provides the back gate electrode. It was further etched into Hall bar geometry as sketched in the inset of Fig.~\ref{Fig1}, and contacted by Cr/Au (5/\NanoMeter{50}) metal leads with a standard edge contact technique~\cite{Wang2013}. Transport measurements were carried out in a He-4 based variable temperature insert cryostat in a temperature range from 1.5 to \SI{100}{\kelvin}. We used standard low-frequency lock-in technique to measure the resistance in  4-terminal geometry with a \NanoAmpere{100} excitation current at a frequency of \SI{12}{\hertz}. 
A gate voltage in the range of $\pm \SI3{\volt}$ was applied through an RC-filter. More details on the transport data are given in the Suppl. Materials.

The photocurrents in the sample were driven by the in-plane alternating electric field $\bm{E}(t)$ of  THz radiation generated by a continuous wave molecular gas laser~\cite{Ganichev2005,Ganichev2009,Olbrich2013}. With formic acid as active medium radiation powers up to \SI{20}{\milli\watt} were obtained for the frequency $f=\SI{0.69}{\tera\hertz}$ (photon energy of $\hbar\omega=\SI{2.87}{\milli\electronvolt}$). The laser spot diameter of about \SI{4}{\milli\meter} is substantially larger than the sample size ensuring uniform illumination of the \gls{tblg} structure. The radiation polarization state was controllably varied by means of lambda-half and lambda-quarter retardation plates, which were used to rotate the electric field vector $\bm E$ of linearly polarized radiation and to change the radiation helicity, respectively. The functional behavior of the Stokes parameters upon rotation of the waveplates is summarized in  the Suppl. Materials, see also Ref.~[\!\!\citenum{Belkov2005}]. The photoresponse was measured as the voltage drop $U_\text{ph}$ directly over the sample resistance applying lock-in technique at a modulation frequency of $\SI{90}{\hertz}$.

\begin{figure*}
	\centering
	\includegraphics[width=\linewidth]{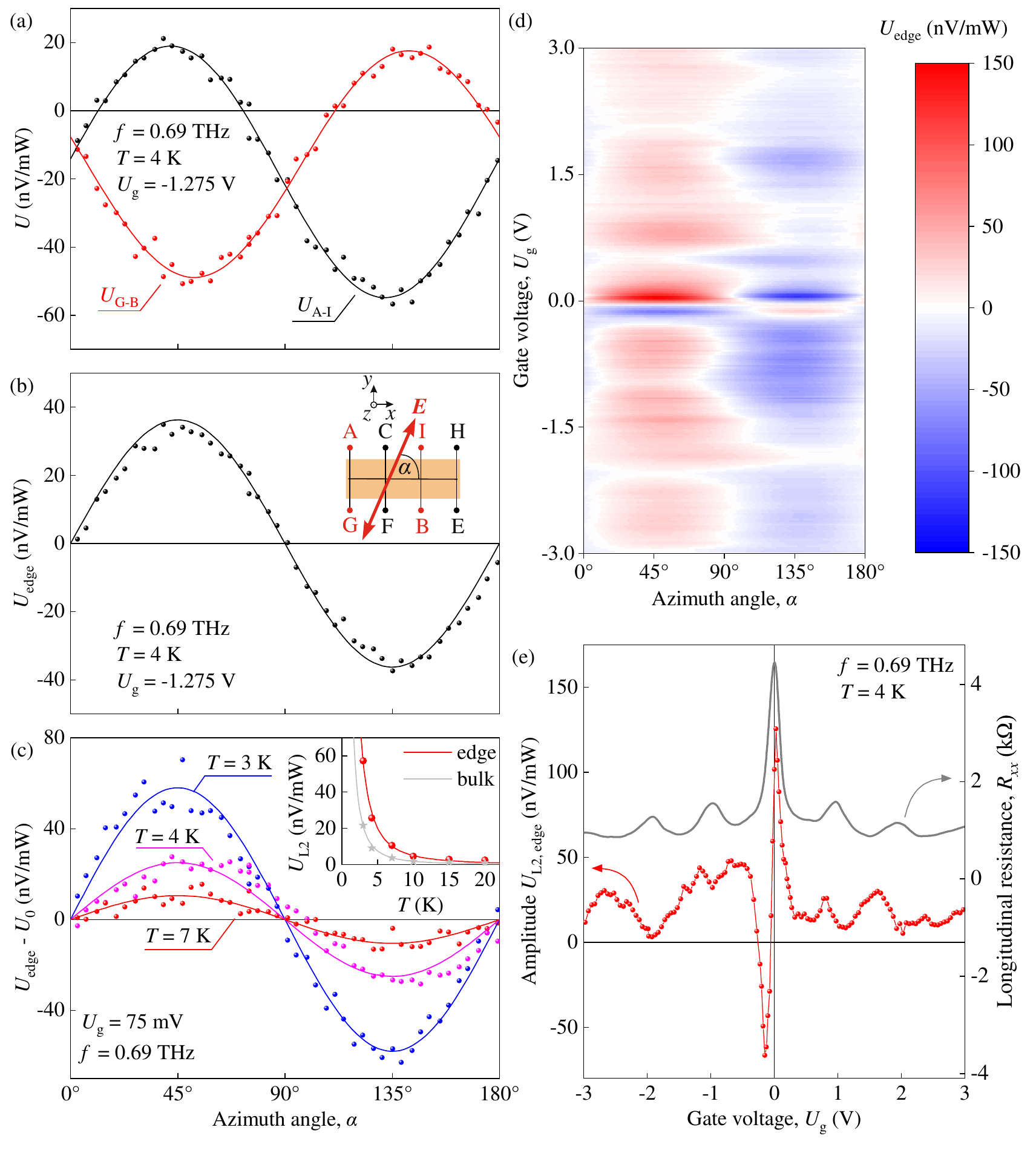}
	\caption{Panel (a): Dependence of the photovoltage on the azimuth angle $\alpha$ measured at edge contacts A-I and G-B (contact arrangement is shown in the inset in panel b) for an applied gate voltage $U_\text{g}=\SI{-1.275}{\volt}$. 		Panel (b): Polarization dependence of the edge photovoltage calculated as the halved difference of the upper two curves. The inset shows the contact geometry and defines angle $\alpha$.  	Panel (c): Dependence of the edge photocurrent on the azimuth angle $\alpha$ for  temperatures of 3, 4, and \SI{7}{\kelvin}. The inset shows the dependence of the parameter $U_\text{L2}$ on the temperature for edge and bulk photocurrents. The data are obtained for $U_\text{g} =\SI{75}{\milli\volt}$. Solid lines represent a guide for the eye.  Panel (d): Color map plot showing the dependence of the edge photocurrent on the gate voltage $U_\text{g}$ and the azimuth angle $\alpha$. Panel (e): Dependence of $U_\text{L2}$ (left axis) on the gate voltage for the edge contribution (red curve). The  gray curve depicts the longitudinal resistance $R_{xx}$.
	}
	\label{Fig2}
\end{figure*}

\textbf{Results}. By illuminating the sample with linearly polarized radiation we observed a photovoltage varying upon rotation of the radiation electric field vector $\bm E$ after
\begin{align}
\label{fit}
U = U_{\text{L}1}\cos 2\alpha  + U_{\text{L}2}\sin 2\alpha+ U_0 \text{ ,}
\end{align}
representing a superposition of the Stokes parameters  $P_{\text{L1}}=\cos 2\alpha$ and $P_{\text{L2}} = \sin 2\alpha$  with different weights given by the fit coefficients $U_{\text{L}1}$, $U_{\text{L}2}$  and $U_{0}$, see Suppl. Materials. Here $\alpha$ is the azimuth angle defining rotation of the radiation electric field vector $\bm E$ with respect to the $x$-direction chosen parallel to the long side of the Hall bar geometry as sketched in the bottom inset of Fig.~\ref{Fig1}~(a). Characteristic polarization dependencies obtained for several gate voltages are shown in Figs.~\ref{Fig1}~(a) and \ref{Fig2}~(a). Measurements for different pairs of contacts revealed that the response is caused by the photocurrents excited in the bulk as well as at the edges of the \gls{tblg} sample. The characteristic feature of the latter one is that the current along opposite edges flows in opposite directions yielding opposite polarities of the photovoltage, as shown in Fig.~\ref{Fig2}~(a) for  contact pairs A-I and G-B. This feature allows us to extract both, the edge contribution by calculating the halved difference of the photovoltages measured at two opposing edges, $U_{\text{edge}} = (U_{\text{A-I}} - U_{\text{G-B}})/2 $ (see Figs.~\ref{Fig2}~(b) and (c)), and the bulk contribution by taking the halved sum of these signals, see inset in Fig.~\ref{Fig2}~(c). The dominating bulk photoresponse is also obtained by measuring signals across the sample as shown for contacts C-F in Fig.~\ref{Fig1}~(a). Note that while in the bulk photocurrent we detected comparable contributions of the terms proportional to $P_{\text{L1}}$ and $P_{\text{L2}}$, in the edge photocurrent the former is zero. Measuring the gate-voltage dependence of the photoresponse we observed that both bulk and edge contributions exhibit oscillations upon variation of $U_\text{g}$ depicted in Figs.~\ref{Fig1}~(b) and \ref{Fig2}~(d,e). These figures point out that the data correlate well with the oscillations of the sheet resistance shown as gray solid curves. Note that in the vicinity of the \gls{cnp}, the photoresponse changes sign and vanishes at gate voltages corresponding to the maximum of $R_{xx}$. Oscillations are clearly detected for all amplitudes ($U_{\text{L}1}$, $U_{\text{L}2}$  and $U_{0}$) in Eq.~\eqref{fit}. An increase of the temperature results in a decrease of the oscillation amplitudes for both sheet resistance, see Suppl. Mater., and the photoresponse. Moreover, for a fixed gate voltage the magnitude of the photocurrent substantially decreases with the temperature increase, see the inset in Fig.~\ref{Fig2}~(c).

\begin{figure}
	\centering
	\includegraphics[width=\linewidth]{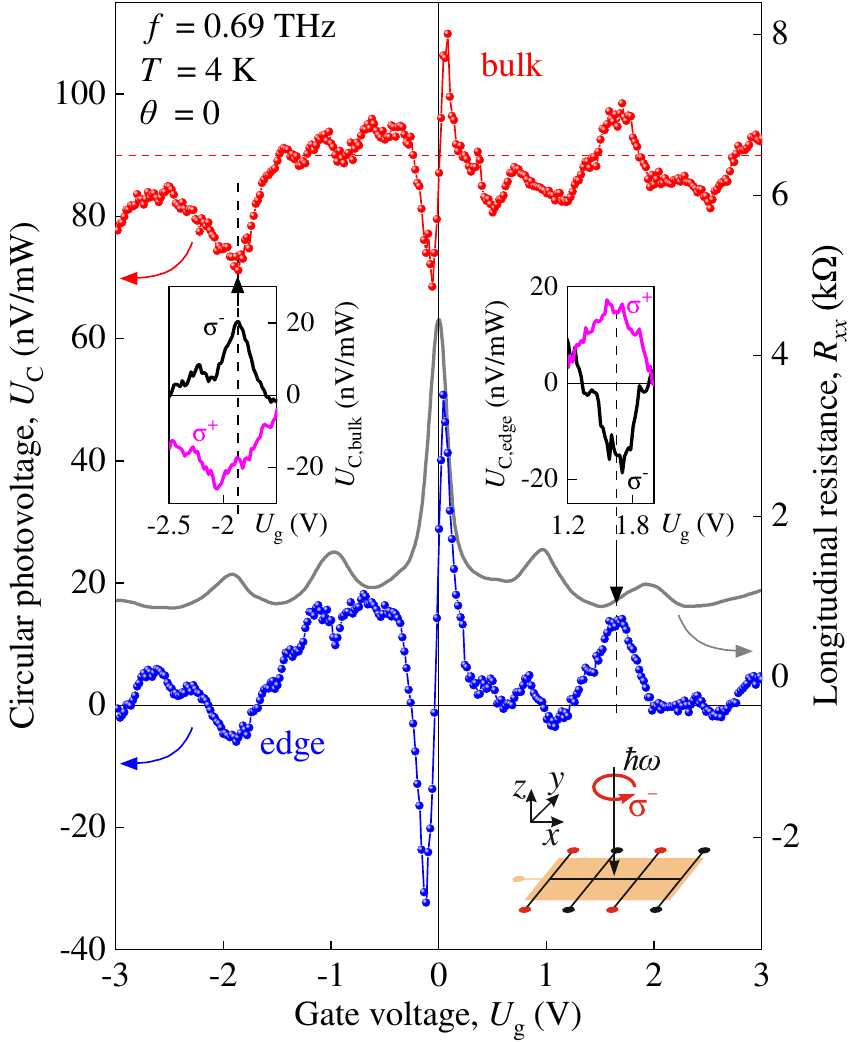}
	\caption{Gate voltage dependencies of the bulk and edge circular photovoltages $U_\text{C} = [ U(\sigma^+) - U(\sigma^-) ]/2$  (left axis).  Note that the bulk photovoltage was offset by 90 nV/mW for visibility.  The gray curve depicts the longitudinal resistance $R_{xx}$. Insets in the middle show dependencies of the bulk and edge photocurrents excited by right- and left-circularly polarized radiation. Bottom inset shows the experimental geometry. 
	}
	\label{Fig3}
\end{figure}

Applying circularly polarized radiation we additionally observed bulk and edge photocurrents, whose direction reverses upon switching the circular polarization from right- ($\sigma^+$) to left-handed ($\sigma^-$) one. This is exemplarily shown in the insets of Fig.~\ref{Fig3} for a certain range of gate voltages. This makes it possible to extract the circular contribution as $U_\text{C}=[U(\sigma^+)-U(\sigma^-)]/2$. Figure~\ref{Fig3} points out that both bulk and edge circular photocurrents, alike earlier discussed photocurrents excited by linearly polarized radiation, oscillate with the variation of the gate voltage and the oscillations correlate with that of the sheet resistance.

\textbf{Theory and discussion.} Stacking the two slightly rotated graphene sheets on top of one another creates a periodic superpotential, the so called moiré potential. While for larger twist-angles the moiré unit cell can be approximated by a hexagonal unit cell and is described by the point group $D_6$, for twist-angles smaller than $1^\circ$, strong lattice reconstruction due to corrugations and strain was observed for \gls{tblg}~\cite{Kim2017a}. These result in a formation of a triangular network of AB and BA regions, which are connected with each other by domain walls with topological character. Since the resulting strain is not uniform in the system, the shape of triangles is distorted and the symmetry of such \gls{tblg} is reduced. Note that the symmetry reduction is also observed in twisted bilayer structures of transition metal dichalcogenides with small twist angles~\cite{Weston2020}. In the samples studied in the current work the twist anlge is approximately $0.6^\circ$. Below we show that our experiments, in particular the presence of the circular photocurrent, give evidence that the symmetry of such \gls{tblg} is reduced to the point group C$_1$. In such systems, excitation with homogeneous radiation at normal incidence results in a photogalvanic current~\cite{Sturman1992,Ivchenko2005}, which is described by
\begin{align}
\label{phenom}
j_i = (C_i P_{\text{L1}}  + S_i P_{\text{L2}} + D_i + \gamma_i P_{\text{circ}}) |\bm E_0|^2. 
\end{align}
Here $i$ is either $x$ or $y$, the \gls{thz} radiation electric field $\bm E = \bm E_0\exp{(-i\omega t)} + \text{c.c.}$,   and $C_{x,y}$, $S_{x,y}$, $D_{x,y}$ and $\gamma_{x,y}$ are eight linearly-independent coefficients which, due to the absence of any nontrivial symmetry operation, are not related to one another. Parameters $P_{\text{L1}}$, $P_{\text{L2}}$ and $P_{\text{circ}}$  are the Stokes parameters. Their variation upon rotation of $\lambda/2$ and $\lambda/4$ plates used in experiments is given in the Suppl. Materials. 

The  photocurrent described by Eq.~\ref{phenom} is in full agreement with experimental results demonstrating the above polarization dependence for different pairs of contacts and revealing comparable values of the coefficients $C_{x,y}$, $S_{x,y}$,  $D_{x,y}$ and $\gamma_{x,y}$, see Figs.~\ref{Fig1}~(a), \ref{Fig2}~(a) and \ref{Fig3}.  Importantly, our observations give evidence for the symmetry reduction to C$_1$ point group in \gls{tblg} with small twist angles, previously suggested on the basis of TEM experiments~\cite{Kim2017a}.  Indeed only this symmetry allows for the polarization independent \gls{lpge} current given by the coefficients $D_{x,y}$ and the helicity driven \gls{cpge}  ($\propto \gamma_{x,y} $), both observed in the experiment.

Now we turn to the microscopic mechanism of the bulk photogalvanic current. In the \gls{thz} range, the electron transport is conveniently described in terms of the electron distribution function $f_{\bm p}$ where $\bm p$ is the two-dimensional electron momentum. 
The distribution function obeys the Boltzmann kinetic equation:
\begin{equation}
	\label{kin_eq}
	{\partial f_{\bm p}\over \partial t} + e\bm E \cdot {\partial f_{\bm p}\over \partial \bm p} = \sum_{\bm p'} (W_{\bm p \bm p'}f_{\bm p'}-W_{\bm p' \bm p}f_{\bm p}).
\end{equation}
Here $W_{\bm p' \bm p}$ is the probability of electron scattering from a state with momentum $\bm p$ to a state with momentum $\bm p'$. The electric current density $\bm j$ is calculated as $	\bm j = e\sum_{\nu,\bm p}f_{\bm p} \bm v_{\bm p}$, where $\nu$ enumerates degenerate states at the Fermi level (spin, valley), and $\bm v_{\bm p}=\partial \varepsilon/\partial \bm p$ is the electron velocity with $\varepsilon(\bm p)$ being the electron dispersion. 

In this kinetic approach, the microscopic reason for the PGE current formation is the electron scattering asymmetry. It is crucial that in non-centrosymmetric systems the scattering probability fulfills $W_{\bm p \bm p'} \neq W_{\bm p' \bm p}$ and can be presented as~\cite{Belinicher1980,Sturman1992,Olbrich2014} $W_{\bm p \bm p'}= W_{\bm p \bm p'}^{s} + W_{\bm p \bm p'}^{a}$, where $W_{\bm p \bm p'}^{s} =W_{\bm p' \bm p}^{s} $ is the symmetric part, and the scattering asymmetry is described by $W_{\bm p \bm p'}^{a} = -W_{\bm p' \bm p}^{a} $, which is nonzero in systems of $C_1$ symmetry.

The photocurrent is a second-order dc response, and the corresponding correction to the distribution function is obtained by two iterations of the kinetic equation in powers of the electric field $\bm E$. The first iteration yields the ac correction $f_{\bm p}^{(1)} = e\tau_{tr} (-f_0') \bm E \cdot \bm v_{\bm p}$ where $f_0(\varepsilon)$ is the Fermi-Dirac equilibrium distribution, prime denotes the derivative with respect to the electron energy, and $\tau_{tr}$ is the transport scattering time. The second iteration is more subtle: there is a dc correction $f_{\bm p}^{(2)}$ obtained accounting for the electric field at the second stage and then for the scattering asymmetry. There is also another dc correction $\delta f_{\bm p}^{(2)}$ which is obtained by the opposite order of perturbations: first accounting for $W_{\bm p \bm p'}^{a}$ and then for the electric field (see Supplementary materials to Ref.~[\!\!\citenum{Olbrich2014}]). 

First we consider the photocurrent driven by linearly polarized radiation. Application of the linearly polarized \gls{thz} field results in an alignment of electron momenta: the corresponding stationary correction to the electron distribution function $f_{\bm p}^{(2)} \propto |\bm E_0|^2\exp{(\pm 2i\varphi_{\bm p})}$  discussed above has a symmetry of the second angular harmonics. The alignment of the electron momenta itself, however, does not lead to the dc electric current. The mechanism of the photocurrent formation is {based on} the scattering asymmetry, which can be visualized as asymmetric elastic scattering on equally oriented triangles. This model was previously developed to describe the photocurrent generation in trigonal 3D and 2D systems~\cite{Belinicher1980,Weber2008,Olbrich2014}. While the model is developed for ideal triangles, it also describes triangle shaped scatterers with distortions. However, in contrast to trigonal systems characterized by  constraints on the coefficients  in Eq.~\eqref{phenom} ($D_x=D_y=C_y=S_x=0$, $C_x=S_y$, where $y$ is parallel to the triangle base), in low symmetric \gls{tblg} with small twist angles these coefficients become linearly independent.

Calculating the photocurrent with the distribution ${f_{\bm p}^{(2)}+\delta f_{\bm p}^{(2)}}$ we obtain that, for linearly-polarized radiation,  it is given by
\begin{align}
	\label{j_fin}
&	j_{x,y} = |\bm E_0|^2 ev_\text{F}\sigma(\omega) \\
&\times \biggl\{
	-{\left[v_\text{F}^3 \tau_2 (\Xi_{c,s}P_{\text{L1}}+\Lambda_{c,s}P_{\text{L2}})\right]'\over v_\text{F}^3} 
+ {1-\omega^2\tau_{tr}\tau_2\over 1+\omega^2\tau_2^2} \nonumber \\
	&\times {\tau_2\over \tau_{tr}} \tau_{tr}' 	[(\Xi_{c,s}\mp\Lambda_{s,c})P_{\text{L1}} \mp (\Xi_{s,c}\pm\Lambda_{c,s})P_{\text{L2}}]
	\biggr\}. \nonumber
\end{align}
Here the prime denotes the derivative with respect to the Fermi energy, $v_\text{F}$ is the Fermi velocity, $\sigma(\omega)=\sigma_0/(1+\omega^2\tau_{tr}^2)$ is the high-frequency conductivity with $\sigma_0$ being the dc conductivity, and the time $\tau_2$ is the relaxation time of the alignment of electron momenta.~\cite{Olbrich2014}. We assumed a parabolic energy dispersion with the density of states independent of the Fermi energy.  The asymmetry of the scattering probability present in the $C_1$ point group is taken into account by the factors $\Xi_{c,s}, \Lambda_{c,s} \ll 1$ defined as:
\begin{align}
	\Xi_{c(s)} = \tau_\text{tr} \sum_{\bm p'}\left< \cos{\varphi_{\bm p}} W^{a}_{\bm p' \bm p} \cos(\sin){(2\varphi_{\bm p'})}\right>_{\varphi_{\bm p}}, \nonumber
	\\
	\Lambda_{c(s)} = \tau_\text{tr} \sum_{\bm p'}\left< \sin{\varphi_{\bm p}} W^{a}_{\bm p' \bm p} \cos(\sin){(2\varphi_{\bm p'})}\right>_{\varphi_{\bm p}}, \nonumber
\end{align}
where the brackets denote averaging over the directions of $\bm p$ at the Fermi circle, and the polar angles are reckoned from the $x$ axis.  All contributions with characteristic polarization dependencies $j \propto P_{\text{L1,2}}$ are clearly detected in the experiment, see e.g. Fig.~\ref{Fig1}~(a). The observed decrease of the photocurrent amplitude with increase of temperature, see the inset of Fig.~\ref{Fig2}~(c), is caused by enhancement of the scattering by phonons which suppresses the carrier mobility. 

Now we turn to \gls{cpge} in the bulk. This ``circular'' photocurrent at normal incidence is forbidden by symmetry in trigonal systems, and its appearance in low twist angle \gls{tblg} is caused by the absence of reflection planes in this system. In the kinetic theory, the helicity-dependent photocurrent is obtained by a procedure similar to the one described above for  linearly-polarized radiation. However, the electron momentum alignment is a distribution which is independent of the light helicity. Therefore, in contrast to linear polarization, the helicity-dependent photocurrent in the kinetic theory appears due to the correction $\delta f_{\bm p}^{(2)}$ only. The resulting photocurrent is given by
\begin{align}
\label{j_circ}
	j_{x,y}^\text{circ} =  ev_\text{F}\sigma(\omega)	{\omega \tau_2 \over 1+\omega^2\tau_2^2}\tau_{tr}' \\ \times \left(1+{\tau_2\over \tau_{tr}}  \right)	(\Lambda_{c,s}\mp \Xi_{s,c}) P_\text{circ} |\bm E_0|^2 .\nonumber
\end{align}
This expression is specific for the $C_1$ symmetry of the studied system. For $C_{3v}$ symmetry, both $j_x^\text{circ}$ and $j_y^\text{circ}$ are equal to zero due to the symmetry constraints  ($\Xi_s=\Lambda_c=0$, $\Xi_c=-\Lambda_s$). Absence of reflection planes in the system under study makes the factors $\Lambda_{c,s}$ and $\Xi_{s,c}$ linearly-independent giving rise to a helicity-dependent  photocurrent at normal incidence. In experiments, this photocurrent manifests itself by opposite sign of the photoresponse excited by right- and left-handed circularly polarized radiation, see middle insets in Fig.~\ref{Fig3}. The temperature dependence of the circular photocurrent is similar to that of the \gls{lpge} current, which is discussed above.

Now we discuss the observed edge photocurrent. Our analysis shows that in \gls{tblg} it has the same nature as the edge currents observed recently in bilayer Bernal-stacked graphene~\cite{Candussio2020}. Theory and mechanism of the edge photogalvanic current are described in detail in Refs. [~\!\!\citenum{Sturman1992,Karch2011,Candussio2020,Glazov2014}], therefore, here we only give a brief description of its basic mechanism. The edge photocurrent is generated due to non-specular (e.g. diffusive) electron scattering at the edge. The corresponding photocurrent, flowing along the edge, is excited either by linearly polarized or circularly polarized radiation and is given by
\begin{equation}
J =  -|\bm E_0|^2 {e\tau_{tr} \over m^\ast} \sigma(\omega)	\left( {P_\text{circ} \over \omega} + P_{\text{L2}}  a \tau_{tr}\right).
\end{equation}
Here $m^\ast$ is the effective mass, $a\sim 1$ is a numerical factor~\cite{Karch2011,Glazov2014,Candussio2020}, and the second Stokes parameter $P_{\text{L2}} =(E_x E_y^\ast + E_yE_x^\ast)/|\bm E_0|^2$ is defined in the axes $(x,y)$ perpendicular to and along the edge, respectively. This expression describes all features observed in the experiment: The edge currents are opposite at opposite edges of the sample, and the edge current $J$ varies at linear polarization as $P_{\text{L2}} =\sin{2\alpha}$, see Fig.~\ref{Fig2}~(b) and (c). The temperature dependence of the edge photocurrent is governed by the factor $\sigma(\omega)$ similar to the bulk currents. Therefore its variation with temperature is similar. 

We note that the theoretical model developed above assumes that the photocurrent is generated due to asymmetric (skew) scattering present due to the $C_1$ symmetry of the structure. In addition, the band structure reconstruction due to the symmetry lowering can result in photocurrent contributions  caused by the Berry curvature and side jumps~\cite{Deyo2009,Moore2010}.

\textbf{Photocurrent oscillations.} The observed photocurrent oscillations upon variation of the gate voltage, see Figs.~\ref{Fig1}~(b),~\ref{Fig2}~(d,e) and~\ref{Fig3} are specific for \gls{tblg} with small twist angle and present the most exciting result of the present work. These oscillations are observed for bulk and edge photogalvanic currents and originate from the band reconstruction. For \gls{tblg} near the second magic angle, Hartree-Fock calculations yield a band structure consisting of multiple relatively flat moiré bands, which are energetically separated and overlapping only by semi-metallic Dirac cones between them~\cite{Lu2020}. The flat dispersion of the bands results in sharp changes of the resistance at continuous variation of the Fermi energy resulting in the filling and emptying of these bands, see gray lines in Figs.~\ref{Fig1}~(b),~\ref{Fig2}~(e) and \ref{Fig3} and	Fig.~2 of the Suppl. Materials. In the most recent work of the Barcelona group describing a detailed magnetotransport study in similar structures nine distinct resistance peaks as a function of carrier density were clearly observed~\cite{Lu2020}. Each resistance peak appears at equally spaced carrier densities being multiples of $n_s$ (where $n_s$ is the carrier density needed to fully fill one moiré band with four electrons per four-fold spin/valley degenerate moiré unit cell). Each resistance peak, therefore, marks the point of inter-band transition between the separate moiré bands. Moreover, the Hall measurements show that each resistance peak in the gate dependent resistivity measurement is accompanied by a Hall density reversal~\cite{Kim2017a}. Consequently, the high-frequency conductivity $\sigma(\omega)$ oscillates upon gate voltage variation causing the oscillations of the photocurrent, which in turn are proportional to $\sigma(\omega)$, see Eqs.~\eqref{j_fin} and~\eqref{j_circ}. This mechanism explains the oscillating behavior of the photocurrent with the gate voltage variation demonstrated in Fig.~\ref{Fig3}. A sign change of all photocurrent contributions detected in the vicinity of the \gls{cnp}, where the resistance exhibits a strong maximum (see Figs.~\ref{Fig1}~(b), \ref{Fig2}~(e) and \ref{Fig3}), is caused by the sign change of the carrier charge. This is a direct consequence of the PGE current, which, in the absence of an external magnetic field, is described by an odd function of the carrier charge (photogalvanics have C-asymmetry). 

The observed temperature smearing of the oscillations in the gate voltage dependence of the photocurrent is caused by redistribution of carriers between the energy bands. A presence of charge carriers in all bands at higher temperatures results in the smoothing of oscillations in the conductivity and, hence, in the photocurrent as seen in Eq.~\eqref{j_fin} where the current is proportional to $\sigma(\omega)$.

\textbf{Summary.} To conclude, we have demonstrated the emergence of terahertz radiation driven photocurrents in tBLG structures which show oscillations as a function of the gate voltage. Our results provide an evidence for the reduction of the point group symmetry of \gls{tblg} with small twist angles to the lowest symmetry group C$_1$.  The oscillations were found to have a common origin with that of the structure resistance: Both are caused by inter-band transition points between the separate almost flat moiré bands. The observed bulk and edge photocurrents may pave the way for a deeper understanding of the rich spectra of nonequilibrium phenomena in \gls{tblg} offering an opto-electronic access for their studies.

\section{Supporting Information}  

\textbf{Laser and methods. }The photocurrents in the sample were driven by the in-plane alternating electric field $\bm{E}(t)$ of  THz radiation generated by a continuous wave molecular gas laser~\cite{Ganichev2005,Ganichev2009,Olbrich2013} operating at wavelength $\lambda=\SI{432}{\micro\meter}$ (corresponding  radiation frequency $f=\SI{0.69}{\tera\hertz}$ and photon energy of $\hbar\omega=\SI{2.87}{\milli\electronvolt}$). The laser beam was focused onto the sample by an off-axis parabolic mirror. The beam profile, monitored by a pyroelectric camera~\cite{Ziemann2000}, has a shape close to the Gaussian fundamental mode. The laser beam was mechanically modulated at a frequency $f_\text{chop}=\SI{90}{\hertz}$ and the phase-locked signal was picked up with lock-in amplifiers. Optical experiments were performed in a helium flow cryostat with $z$-cut crystalline quartz windows allowing the coupling of the sample with normally incident polarized \gls{thz} radiation.  Additionally, the entrance window was covered by a thin black polyethylene foil preventing sample illumination by visible and near-infrared light.
%
%
\begin{figure*}
	\centering
	\includegraphics[width=\linewidth]{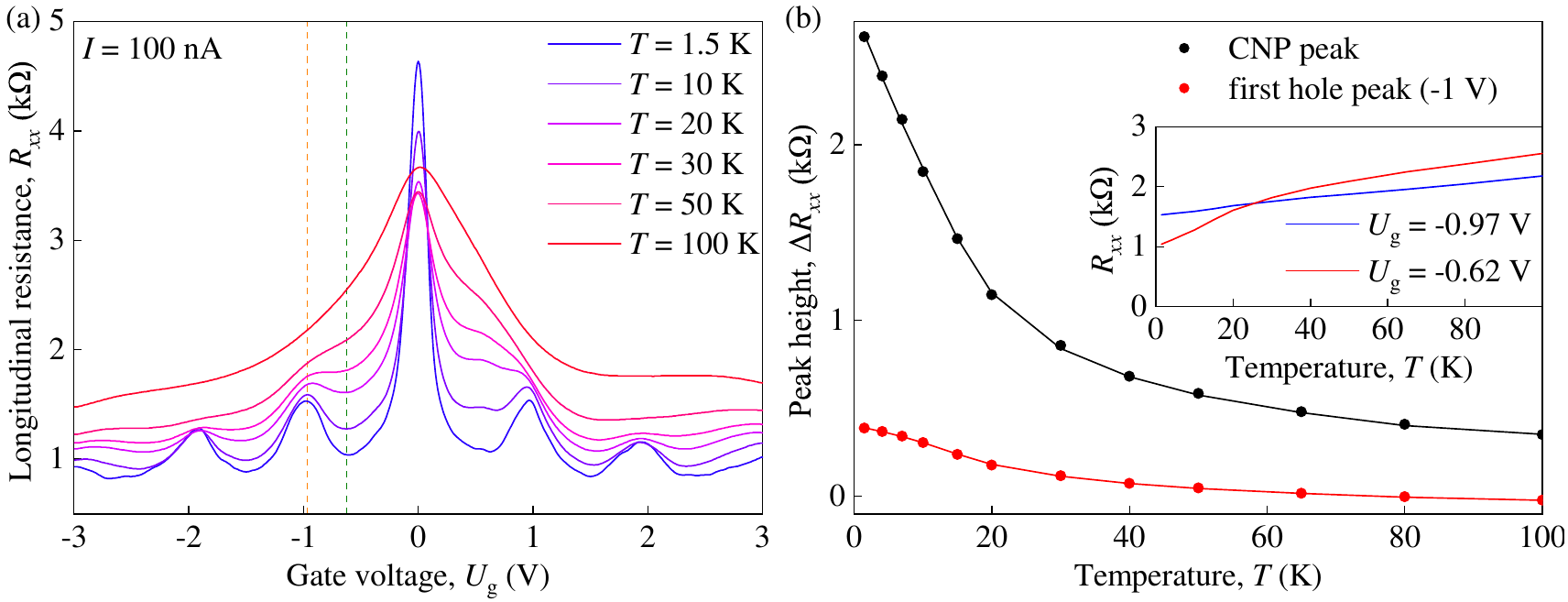}
	\caption{Panel (a): Dependence of the longitudinal resistance $R_{xx}$ on the gate voltage $U_\text{g}$ for various temperatures between 1.5 and \SI{100}{\kelvin}. Panel (b): Dependence of the peak height for the resistance peaks at the charge neutrality point at $U_\text{g}=0$ (black curve) and the first oscillation peak at $U_\text{g}\approx \SI{-1}{\volt}$ as a function of the temperature. 	The peak height  was calculated according to $\Delta R_{xx}=R_{xx}-\braket{R_{xx}}$, where $\braket{R_{xx}}$ denotes the non-oscillating background obtained by a 65-point FFT smooth.}
	\label{Fig_transport}
\end{figure*}

\textbf{Polarization measurements and variation of the Stokes parameters.} The developed theory and experiment show that the total photocurrent is proportional to the combination of the Stokes parameters with different weights. The Stokes parameters include the degrees of linear polarization $P_{\text{L1}}$ in the basis ($x$, $y$), and  $P_{\text{L2}}$ in the basis ($\tilde x$, $\tilde y$) rotated by \Degree{45},  as well as the radiation helicity $P_{\text{circ}}$ and parameter describing the radiation intensity $s_0$. These parameters are given by
\begin{align}
\label{phenomStokes}
&P_{\text{L1}}=\frac{|E_x|^2-|E_y|^2}{|\bm E_0|^2} \,, \nonumber\\  
& P_{\text{L2}}=\frac{E_x E_y^* + E_yE_x^*}{|\bm E_0|^2} \, , \nonumber\\ 
&P_{\text{circ}} = \frac{i(E_x E_y^* -E_yE_x^*)}{|\bm E_0|^2}=\frac{I^{\sigma^+}-I^{\sigma^-}}{I^{\sigma^+}+I^{\sigma^-}}, \nonumber\\  
&s_0 = |E_x|^2+|E_y|^2 \,.
\end{align}
with the intensity $I^{\sigma^+}$ ($I^{\sigma^-}$) of the right- {(left-)} handed polarized component, see e.g., Ref.~[\!\!\citenum{Saleh2007}].

In the main text of this work we focused on the dependencies of the photoresponse on the azimuthal angle $\alpha$ and the difference between  polarity and magnitude of the photocurrent excited by right- and left-handed circularly polarized radiation. To vary the radiation polarization state we used room-temperature crystal quartz lambda-half and lambda-quarter retardation plates.  In the former case rotation of the lambda-half plate results in the change of the first two Stokes parameters as following:
$$P_{\text{L1}}(\alpha)=\cos2\alpha$$ 
and 
$$P_{\text{L2}}(\alpha)=\sin2\alpha,$$
where $\alpha$ is the azimuth angle defining rotation of the radiation electric field vector $\bm E$ with respect to the $x$-direction. In the latter case, for the angle $\varphi$ between initially linearly polarized laser radiation and the optical axis (fast direction) of the plate being equal to  $\varphi = 45^\circ$ and $135^\circ$ we obtained right- and left-handed circularly polarized radiation. Under these conditions the parameters $P_{\text{L1}}(\varphi)$ and $P_{\text{L2}}(\varphi)$ become zero, and the circular photocurrent can be calculated as a difference of the photoresponses obtained for $\varphi = 45^\circ$ and $\varphi = 135^\circ$. Note that for this setup the Stokes parameters vary with the angle $\varphi$ as~\cite{Belkov2005,Weber2008} :
$$P_{\text{L1}}(\varphi) = (\cos 4\varphi + 1)/2,$$
$$P_{\text{L2}}(\varphi) = \sin (4\varphi)/2$$
and
$$P_{\rm circ} = \sin 2\varphi\:.$$



\textbf{Magnetotransport experiments.} Figure~\ref{Fig_transport}~(a) shows the gate voltage dependence of the twisted bilayer graphene resistance obtained at different temperatures. To continuously tune the Fermi energy in our system we apply a gate voltage $U_\text{g}$ to the graphite back gate. Note that for different sample cooldowns, the \gls{cnp} can occur at slightly different gate voltages.  This is caused by, e.g.,  cooldown dependent charge trapping in the insulator. The  low temperature data clearly show oscillations of the longitudinal resistance $R_{xx}$, which vanish with the increase of temperature. The temperature dependence of the resistance peak height is shown in  Fig.~\ref{Fig_transport}(b) .

\textbf{Acknowledgements.} The support from the Deutsche Forschungsgemeinschaft (DFG, German Research Foundation) - Project SPP 2244 (GA501/17-1),  the Volkswagen Stiftung Program (97738) and the IRAP program of the Foundation for Polish Science (grant MAB/2018/9, project CENTERA) is gratefully acknowledged. L. E. G. thanks the  financial support of the Russian Science Foundation (Project No. 17-12-01265) and the Foundation for Advancement of Theoretical Physics and Mathematics (``BASIS'').

\textbf{Notes.} The authors declare no competing financial interest.

\bibliography{twisted_paper_shortened_sdg4}

\providecommand{\latin}[1]{#1}
\makeatletter
\providecommand{\doi}
  {\begingroup\let\do\@makeother\dospecials
  \catcode`\{=1 \catcode`\}=2 \doi@aux}
\providecommand{\doi@aux}[1]{\endgroup\texttt{#1}}
\makeatother
\providecommand*\mcitethebibliography{\thebibliography}
\csname @ifundefined\endcsname{endmcitethebibliography}
  {\let\endmcitethebibliography\endthebibliography}{}
\begin{mcitethebibliography}{45}
\providecommand*\natexlab[1]{#1}
\providecommand*\mciteSetBstSublistMode[1]{}
\providecommand*\mciteSetBstMaxWidthForm[2]{}
\providecommand*\mciteBstWouldAddEndPuncttrue
  {\def\EndOfBibitem{\unskip.}}
\providecommand*\mciteBstWouldAddEndPunctfalse
  {\let\EndOfBibitem\relax}
\providecommand*\mciteSetBstMidEndSepPunct[3]{}
\providecommand*\mciteSetBstSublistLabelBeginEnd[3]{}
\providecommand*\EndOfBibitem{}
\mciteSetBstSublistMode{f}
\mciteSetBstMaxWidthForm{subitem}{(\alph{mcitesubitemcount})}
\mciteSetBstSublistLabelBeginEnd
  {\mcitemaxwidthsubitemform\space}
  {\relax}
  {\relax}

\bibitem[dos Santos \latin{et~al.}(2007)dos Santos, Peres, and
  Neto]{Santos2007}
dos Santos,~J. M. B.~L.; Peres,~N. M.~R.; Neto,~A. H.~C. Graphene Bilayer with
  a Twist: Electronic Structure. \emph{Phys. Rev. Lett.} \textbf{2007},
  \emph{99}, 256802\relax
\mciteBstWouldAddEndPuncttrue
\mciteSetBstMidEndSepPunct{\mcitedefaultmidpunct}
{\mcitedefaultendpunct}{\mcitedefaultseppunct}\relax
\EndOfBibitem
\bibitem[Li \latin{et~al.}(2009)Li, Luican, dos Santos, Neto, Reina, Kong, and
  Andrei]{Li2009}
Li,~G.; Luican,~A.; dos Santos,~J. M. B.~L.; Neto,~A. H.~C.; Reina,~A.;
  Kong,~J.; Andrei,~E.~Y. Observation of Van Hove singularities in twisted
  graphene layers. \emph{Nat. Phys.} \textbf{2009}, \emph{6}, 109\relax
\mciteBstWouldAddEndPuncttrue
\mciteSetBstMidEndSepPunct{\mcitedefaultmidpunct}
{\mcitedefaultendpunct}{\mcitedefaultseppunct}\relax
\EndOfBibitem
\bibitem[Mele(2010)]{Mele2010}
Mele,~E.~J. Commensuration and interlayer coherence in twisted bilayer
  graphene. \emph{Phys. Rev. B} \textbf{2010}, \emph{81}, 161405\relax
\mciteBstWouldAddEndPuncttrue
\mciteSetBstMidEndSepPunct{\mcitedefaultmidpunct}
{\mcitedefaultendpunct}{\mcitedefaultseppunct}\relax
\EndOfBibitem
\bibitem[Li \latin{et~al.}(2010)Li, Wang, Zhu, Liu, Ye, Chen, Wang, He, Wang,
  Ma, Zhang, Dai, Fang, Xie, Liu, Qi, Jia, Zhang, and Xue]{Li2010}
Li,~Y.-Y. \latin{et~al.}  Intrinsic Topological Insulator Bi$_2$Te$_3$ Thin
  Films on Si and Their Thickness Limit. \emph{Advanced Mat.} \textbf{2010},
  \emph{22}, 4002\relax
\mciteBstWouldAddEndPuncttrue
\mciteSetBstMidEndSepPunct{\mcitedefaultmidpunct}
{\mcitedefaultendpunct}{\mcitedefaultseppunct}\relax
\EndOfBibitem
\bibitem[Morell \latin{et~al.}(2010)Morell, Correa, Vargas, Pacheco, and
  Barticevic]{Morell2010}
Morell,~E.~S.; Correa,~J.~D.; Vargas,~P.; Pacheco,~M.; Barticevic,~Z. Flat
  bands in slightly twisted bilayer graphene: Tight-binding calculations.
  \emph{Phys. Rev. B} \textbf{2010}, \emph{82}, 121407\relax
\mciteBstWouldAddEndPuncttrue
\mciteSetBstMidEndSepPunct{\mcitedefaultmidpunct}
{\mcitedefaultendpunct}{\mcitedefaultseppunct}\relax
\EndOfBibitem
\bibitem[Bistritzer and MacDonald(2011)Bistritzer, and
  MacDonald]{Bistritzer2011}
Bistritzer,~R.; MacDonald,~A.~H. Moire bands in twisted double-layer graphene.
  \emph{Proc. Nat. Acad. Sci.} \textbf{2011}, \emph{108}, 12233\relax
\mciteBstWouldAddEndPuncttrue
\mciteSetBstMidEndSepPunct{\mcitedefaultmidpunct}
{\mcitedefaultendpunct}{\mcitedefaultseppunct}\relax
\EndOfBibitem
\bibitem[Kim \latin{et~al.}(2017)Kim, DaSilva, Huang, Fallahazad, Larentis,
  Taniguchi, Watanabe, LeRoy, MacDonald, and Tutuc]{Kim2017a}
Kim,~K.; DaSilva,~A.; Huang,~S.; Fallahazad,~B.; Larentis,~S.; Taniguchi,~T.;
  Watanabe,~K.; LeRoy,~B.~J.; MacDonald,~A.~H.; Tutuc,~E. Tunable moir{\'{e}}
  bands and strong correlations in small-twist-angle bilayer graphene.
  \emph{Proc. Natl. Acad. Sci.} \textbf{2017}, \emph{114}, 3364--3369\relax
\mciteBstWouldAddEndPuncttrue
\mciteSetBstMidEndSepPunct{\mcitedefaultmidpunct}
{\mcitedefaultendpunct}{\mcitedefaultseppunct}\relax
\EndOfBibitem
\bibitem[Cao \latin{et~al.}(2018)Cao, Fatemi, Demir, Fang, Tomarken, Luo,
  Sanchez-Yamagishi, Watanabe, Taniguchi, Kaxiras, Ashoori, and
  Jarillo-Herrero]{Cao2018a}
Cao,~Y.; Fatemi,~V.; Demir,~A.; Fang,~S.; Tomarken,~S.~L.; Luo,~J.~Y.;
  Sanchez-Yamagishi,~J.~D.; Watanabe,~K.; Taniguchi,~T.; Kaxiras,~E.;
  Ashoori,~R.~C.; Jarillo-Herrero,~P. Correlated insulator behaviour at
  half-filling in magic-angle graphene superlattices. \emph{Nature}
  \textbf{2018}, \emph{556}, 80\relax
\mciteBstWouldAddEndPuncttrue
\mciteSetBstMidEndSepPunct{\mcitedefaultmidpunct}
{\mcitedefaultendpunct}{\mcitedefaultseppunct}\relax
\EndOfBibitem
\bibitem[Cao \latin{et~al.}(2018)Cao, Fatemi, Fang, Watanabe, Taniguchi,
  Kaxiras, and Jarillo-Herrero]{Cao2018}
Cao,~Y.; Fatemi,~V.; Fang,~S.; Watanabe,~K.; Taniguchi,~T.; Kaxiras,~E.;
  Jarillo-Herrero,~P. Unconventional superconductivity in magic-angle graphene
  superlattices. \emph{Nature} \textbf{2018}, \emph{556}, 43\relax
\mciteBstWouldAddEndPuncttrue
\mciteSetBstMidEndSepPunct{\mcitedefaultmidpunct}
{\mcitedefaultendpunct}{\mcitedefaultseppunct}\relax
\EndOfBibitem
\bibitem[Sunku \latin{et~al.}(2018)Sunku, Ni, Jiang, Yoo, Sternbach, McLeod,
  Stauber, Xiong, Taniguchi, Watanabe, Kim, Fogler, and Basov]{Sunku2018}
Sunku,~S.~S.; Ni,~G.~X.; Jiang,~B.~Y.; Yoo,~H.; Sternbach,~A.; McLeod,~A.~S.;
  Stauber,~T.; Xiong,~L.; Taniguchi,~T.; Watanabe,~K.; Kim,~P.; Fogler,~M.~M.;
  Basov,~D.~N. Photonic crystals for nano-light in moir{\'{e}} graphene
  superlattices. \emph{Science} \textbf{2018}, \emph{362}, 1153\relax
\mciteBstWouldAddEndPuncttrue
\mciteSetBstMidEndSepPunct{\mcitedefaultmidpunct}
{\mcitedefaultendpunct}{\mcitedefaultseppunct}\relax
\EndOfBibitem
\bibitem[Yoo \latin{et~al.}(2019)Yoo, Engelke, Carr, Fang, Zhang, Cazeaux,
  Sung, Hovden, Tsen, Taniguchi, Watanabe, Yi, Kim, Luskin, Tadmor, Kaxiras,
  and Kim]{Yoo2019}
Yoo,~H. \latin{et~al.}  Atomic and electronic reconstruction at the van der
  Waals interface in twisted bilayer graphene. \emph{Nat. Mater.}
  \textbf{2019}, \emph{18}, 448\relax
\mciteBstWouldAddEndPuncttrue
\mciteSetBstMidEndSepPunct{\mcitedefaultmidpunct}
{\mcitedefaultendpunct}{\mcitedefaultseppunct}\relax
\EndOfBibitem
\bibitem[Jiang \latin{et~al.}(2019)Jiang, Lai, Watanabe, Taniguchi, Haule, Mao,
  and Andrei]{Jiang2019}
Jiang,~Y.; Lai,~X.; Watanabe,~K.; Taniguchi,~T.; Haule,~K.; Mao,~J.;
  Andrei,~E.~Y. Charge order and broken rotational symmetry in magic-angle
  twisted bilayer graphene. \emph{Nature} \textbf{2019}, \emph{573}, 91\relax
\mciteBstWouldAddEndPuncttrue
\mciteSetBstMidEndSepPunct{\mcitedefaultmidpunct}
{\mcitedefaultendpunct}{\mcitedefaultseppunct}\relax
\EndOfBibitem
\bibitem[Yankowitz \latin{et~al.}(2019)Yankowitz, Chen, Polshyn, Zhang,
  Watanabe, Taniguchi, Graf, Young, and Dean]{Yankowitz2019}
Yankowitz,~M.; Chen,~S.; Polshyn,~H.; Zhang,~Y.; Watanabe,~K.; Taniguchi,~T.;
  Graf,~D.; Young,~A.~F.; Dean,~C.~R. Tuning superconductivity in twisted
  bilayer graphene. \emph{Science} \textbf{2019}, \emph{363}, 1059\relax
\mciteBstWouldAddEndPuncttrue
\mciteSetBstMidEndSepPunct{\mcitedefaultmidpunct}
{\mcitedefaultendpunct}{\mcitedefaultseppunct}\relax
\EndOfBibitem
\bibitem[Lu \latin{et~al.}(2019)Lu, Stepanov, Yang, Xie, Aamir, Das, Urgell,
  Watanabe, Taniguchi, Zhang, Bachtold, MacDonald, and Efetov]{Lu2019}
Lu,~X.; Stepanov,~P.; Yang,~W.; Xie,~M.; Aamir,~M.~A.; Das,~I.; Urgell,~C.;
  Watanabe,~K.; Taniguchi,~T.; Zhang,~G.; Bachtold,~A.; MacDonald,~A.~H.;
  Efetov,~D.~K. Superconductors, orbital magnets and correlated states in
  magic-angle bilayer graphene. \emph{Nature} \textbf{2019}, \emph{574},
  653\relax
\mciteBstWouldAddEndPuncttrue
\mciteSetBstMidEndSepPunct{\mcitedefaultmidpunct}
{\mcitedefaultendpunct}{\mcitedefaultseppunct}\relax
\EndOfBibitem
\bibitem[Hesp \latin{et~al.}(2019)Hesp, Torre, Rodan-Legrain, Novelli, Cao,
  Carr, Fang, Stepanov, Barcons-Ruiz, Herzig-Sheinfux, Watanabe, Taniguchi,
  Efetov, Kaxiras, Jarillo-Herrero, Polini, and Koppens]{Hesp2019}
Hesp,~N. C.~H. \latin{et~al.}  Collective excitations in twisted bilayer
  graphene close to the magic angle. \emph{arxiv:1910.07893} \textbf{2019}.
  \relax
\mciteBstWouldAddEndPunctfalse
\mciteSetBstMidEndSepPunct{\mcitedefaultmidpunct}
{}{\mcitedefaultseppunct}\relax
\EndOfBibitem
\bibitem[Glazov and Ganichev(2014)Glazov, and Ganichev]{Glazov2014}
Glazov,~M.; Ganichev,~S. High frequency electric field induced nonlinear
  effects in graphene. \emph{Phys. Rep.} \textbf{2014}, \emph{535}, 101\relax
\mciteBstWouldAddEndPuncttrue
\mciteSetBstMidEndSepPunct{\mcitedefaultmidpunct}
{\mcitedefaultendpunct}{\mcitedefaultseppunct}\relax
\EndOfBibitem
\bibitem[Karch \latin{et~al.}(2010)Karch, Olbrich, Schmalzbauer, Brinsteiner,
  Wurstbauer, Glazov, Tarasenko, Ivchenko, Weiss, Eroms, and
  Ganichev]{Karch2010}
Karch,~J.; Olbrich,~P.; Schmalzbauer,~M.; Brinsteiner,~C.; Wurstbauer,~U.;
  Glazov,~M.~M.; Tarasenko,~S.~A.; Ivchenko,~E.~L.; Weiss,~D.; Eroms,~J.;
  Ganichev,~S.~D. Photon helicity driven electric currents in graphene.
  \emph{arXiv:1002.1047v1} \textbf{2010}. \relax
\mciteBstWouldAddEndPunctfalse
\mciteSetBstMidEndSepPunct{\mcitedefaultmidpunct}
{}{\mcitedefaultseppunct}\relax
\EndOfBibitem
\bibitem[Karch \latin{et~al.}(2011)Karch, Drexler, Olbrich, Fehrenbacher,
  Hirmer, Glazov, Tarasenko, Ivchenko, Birkner, Eroms, Weiss, Yakimova,
  Lara-Avila, Kubatkin, Ostler, Seyller, and Ganichev]{Karch2011}
Karch,~J. \latin{et~al.}  Terahertz Radiation Driven Chiral Edge Currents in
  Graphene. \emph{Phys. Rev. Lett.} \textbf{2011}, \emph{107}, 276601\relax
\mciteBstWouldAddEndPuncttrue
\mciteSetBstMidEndSepPunct{\mcitedefaultmidpunct}
{\mcitedefaultendpunct}{\mcitedefaultseppunct}\relax
\EndOfBibitem
\bibitem[Jiang \latin{et~al.}(2011)Jiang, Shalygin, Panevin, Danilov, Glazov,
  Yakimova, Lara-Avila, Kubatkin, and Ganichev]{Jiang2011}
Jiang,~C.; Shalygin,~V.~A.; Panevin,~V.~Y.; Danilov,~S.~N.; Glazov,~M.~M.;
  Yakimova,~R.; Lara-Avila,~S.; Kubatkin,~S.; Ganichev,~S.~D.
  Helicity-dependent photocurrents in graphene layers excited by midinfrared
  radiation of a CO$_2$ laser. \emph{Phys. Rev. B} \textbf{2011}, \emph{84},
  125429\relax
\mciteBstWouldAddEndPuncttrue
\mciteSetBstMidEndSepPunct{\mcitedefaultmidpunct}
{\mcitedefaultendpunct}{\mcitedefaultseppunct}\relax
\EndOfBibitem
\bibitem[Olbrich \latin{et~al.}(2013)Olbrich, Drexler, Golub, Danilov,
  Shalygin, Yakimova, Lara-Avila, Kubatkin, Redlich, Huber, and
  Ganichev]{Olbrich2013a}
Olbrich,~P.; Drexler,~C.; Golub,~L.~E.; Danilov,~S.~N.; Shalygin,~V.~A.;
  Yakimova,~R.; Lara-Avila,~S.; Kubatkin,~S.; Redlich,~B.; Huber,~R.;
  Ganichev,~S.~D. Reststrahl band-assisted photocurrents in epitaxial graphene
  layers. \emph{Phys. Rev. B} \textbf{2013}, \emph{88}, 245425\relax
\mciteBstWouldAddEndPuncttrue
\mciteSetBstMidEndSepPunct{\mcitedefaultmidpunct}
{\mcitedefaultendpunct}{\mcitedefaultseppunct}\relax
\EndOfBibitem
\bibitem[Maysonnave \latin{et~al.}(2014)Maysonnave, Huppert, Wang, Maero,
  Berger, de~Heer, Norris, Vaulchier, Dhillon, Tignon, Ferreira, and
  Mangeney]{Maysonnave2014}
Maysonnave,~J.; Huppert,~S.; Wang,~F.; Maero,~S.; Berger,~C.; de~Heer,~W.;
  Norris,~T.~B.; Vaulchier,~L. A.~D.; Dhillon,~S.; Tignon,~J.; Ferreira,~R.;
  Mangeney,~J. Terahertz Generation by Dynamical Photon Drag Effect in Graphene
  Excited by Femtosecond Optical Pulses. \emph{Nano Lett.} \textbf{2014},
  \emph{14}, 5797\relax
\mciteBstWouldAddEndPuncttrue
\mciteSetBstMidEndSepPunct{\mcitedefaultmidpunct}
{\mcitedefaultendpunct}{\mcitedefaultseppunct}\relax
\EndOfBibitem
\bibitem[Obraztsov \latin{et~al.}(2014)Obraztsov, Kanda, Konishi,
  Kuwata-Gonokami, Garnov, Obraztsov, and Svirko]{Obraztsov2014}
Obraztsov,~P.~A.; Kanda,~N.; Konishi,~K.; Kuwata-Gonokami,~M.; Garnov,~S.~V.;
  Obraztsov,~A.~N.; Svirko,~Y.~P. Photon-drag-induced terahertz emission from
  graphene. \emph{Phys. Rev. B} \textbf{2014}, \emph{90}, 241416\relax
\mciteBstWouldAddEndPuncttrue
\mciteSetBstMidEndSepPunct{\mcitedefaultmidpunct}
{\mcitedefaultendpunct}{\mcitedefaultseppunct}\relax
\EndOfBibitem
\bibitem[Inglot \latin{et~al.}(2015)Inglot, Dugaev, Sherman, and
  Barna{\'{s}}]{Inglot2015}
Inglot,~M.; Dugaev,~V.~K.; Sherman,~E.~Y.; Barna{\'{s}},~J. Enhanced
  photogalvanic effect in graphene due to Rashba spin-orbit coupling.
  \emph{Phys. Rev. B} \textbf{2015}, \emph{91}, 195428\relax
\mciteBstWouldAddEndPuncttrue
\mciteSetBstMidEndSepPunct{\mcitedefaultmidpunct}
{\mcitedefaultendpunct}{\mcitedefaultseppunct}\relax
\EndOfBibitem
\bibitem[Hipolito \latin{et~al.}(2016)Hipolito, Pedersen, and
  Pereira]{Hipolito2016}
Hipolito,~F.; Pedersen,~T.~G.; Pereira,~V.~M. Nonlinear photocurrents in
  two-dimensional systems based on graphene and boron nitride. \emph{Phys. Rev.
  B} \textbf{2016}, \emph{94}, 045434\relax
\mciteBstWouldAddEndPuncttrue
\mciteSetBstMidEndSepPunct{\mcitedefaultmidpunct}
{\mcitedefaultendpunct}{\mcitedefaultseppunct}\relax
\EndOfBibitem
\bibitem[Ganichev \latin{et~al.}(2017)Ganichev, Weiss, and Eroms]{Ganichev2017}
Ganichev,~S.~D.; Weiss,~D.; Eroms,~J. Terahertz Electric Field Driven Electric
  Currents and Ratchet Effects in Graphene. \emph{Ann. Phys.} \textbf{2017},
  \emph{529}, 1600406\relax
\mciteBstWouldAddEndPuncttrue
\mciteSetBstMidEndSepPunct{\mcitedefaultmidpunct}
{\mcitedefaultendpunct}{\mcitedefaultseppunct}\relax
\EndOfBibitem
\bibitem[Zhu \latin{et~al.}(2017)Zhu, Huang, Yao, Quan, Zhang, Li, Gu, Xu, and
  Ren]{Zhu2017}
Zhu,~L.; Huang,~Y.; Yao,~Z.; Quan,~B.; Zhang,~L.; Li,~J.; Gu,~C.; Xu,~X.;
  Ren,~Z. Enhanced polarization-sensitive terahertz emission from vertically
  grown graphene by a dynamical photon drag effect. \emph{Nanoscale}
  \textbf{2017}, \emph{9}, 10301\relax
\mciteBstWouldAddEndPuncttrue
\mciteSetBstMidEndSepPunct{\mcitedefaultmidpunct}
{\mcitedefaultendpunct}{\mcitedefaultseppunct}\relax
\EndOfBibitem
\bibitem[Plank \latin{et~al.}(2018)Plank, Durnev, Candussio, Pernul, Dantscher,
  Mönch, Sandner, Eroms, Weiss, Bel'kov, Tarasenko, and Ganichev]{Plank2018a}
Plank,~H.; Durnev,~M.~V.; Candussio,~S.; Pernul,~J.; Dantscher,~K.-M.;
  Mönch,~E.; Sandner,~A.; Eroms,~J.; Weiss,~D.; Bel'kov,~V.~V.;
  Tarasenko,~S.~A.; Ganichev,~S.~D. Edge currents driven by terahertz radiation
  in graphene in quantum Hall regime. \emph{2D Mater.} \textbf{2018}, \emph{6},
  011002\relax
\mciteBstWouldAddEndPuncttrue
\mciteSetBstMidEndSepPunct{\mcitedefaultmidpunct}
{\mcitedefaultendpunct}{\mcitedefaultseppunct}\relax
\EndOfBibitem
\bibitem[Candussio \latin{et~al.}(2020)Candussio, Durnev, Tarasenko, Yin, Keil,
  Yang, Son, Mishchenko, Plank, Bel'kov, Slizovskiy, Fal'ko, and
  Ganichev]{Candussio2020}
Candussio,~S.; Durnev,~M.; Tarasenko,~S.; Yin,~J.; Keil,~J.; Yang,~Y.;
  Son,~S.-K.; Mishchenko,~A.; Plank,~H.; Bel'kov,~V.; Slizovskiy,~S.;
  Fal'ko,~V.; Ganichev,~S. Edge photocurrent driven by THz electric field in
  bi-layer graphene. \emph{arxiv:2005.01407} \textbf{2020}. \relax
\mciteBstWouldAddEndPunctfalse
\mciteSetBstMidEndSepPunct{\mcitedefaultmidpunct}
{}{\mcitedefaultseppunct}\relax
\EndOfBibitem
\bibitem[Wang \latin{et~al.}(2013)Wang, Meric, Huang, Gao, Gao, Tran,
  Taniguchi, Watanabe, Campos, Muller, Guo, Kim, Hone, Shepard, and
  Dean]{Wang2013}
Wang,~L.; Meric,~I.; Huang,~P.~Y.; Gao,~Q.; Gao,~Y.; Tran,~H.; Taniguchi,~T.;
  Watanabe,~K.; Campos,~L.~M.; Muller,~D.~A.; Guo,~J.; Kim,~P.; Hone,~J.;
  Shepard,~K.~L.; Dean,~C.~R. One-Dimensional Electrical Contact to a
  Two-Dimensional Material. \emph{Science} \textbf{2013}, \emph{342}, 614\relax
\mciteBstWouldAddEndPuncttrue
\mciteSetBstMidEndSepPunct{\mcitedefaultmidpunct}
{\mcitedefaultendpunct}{\mcitedefaultseppunct}\relax
\EndOfBibitem
\bibitem[Ganichev and Prettl(2005)Ganichev, and Prettl]{Ganichev2005}
Ganichev,~S.~D.; Prettl,~W. \emph{Intense Terahertz Excitation of
  Semiconductors}; Oxford University Press: Oxford, 2005\relax
\mciteBstWouldAddEndPuncttrue
\mciteSetBstMidEndSepPunct{\mcitedefaultmidpunct}
{\mcitedefaultendpunct}{\mcitedefaultseppunct}\relax
\EndOfBibitem
\bibitem[Ganichev \latin{et~al.}(2009)Ganichev, Tarasenko, Bel'kov, Olbrich,
  Eder, Yakovlev, Kolkovsky, Zaleszczyk, Karczewski, Wojtowicz, and
  Weiss]{Ganichev2009}
Ganichev,~S.~D.; Tarasenko,~S.~A.; Bel'kov,~V.~V.; Olbrich,~P.; Eder,~W.;
  Yakovlev,~D.~R.; Kolkovsky,~V.; Zaleszczyk,~W.; Karczewski,~G.;
  Wojtowicz,~T.; Weiss,~D. Spin Currents in Diluted Magnetic Semiconductors.
  \emph{Phys. Rev. Lett.} \textbf{2009}, \emph{102}, 156602\relax
\mciteBstWouldAddEndPuncttrue
\mciteSetBstMidEndSepPunct{\mcitedefaultmidpunct}
{\mcitedefaultendpunct}{\mcitedefaultseppunct}\relax
\EndOfBibitem
\bibitem[Olbrich \latin{et~al.}(2013)Olbrich, Zoth, Vierling, Dantscher,
  Budkin, Tarasenko, Bel'kov, Kozlov, Kvon, Mikhailov, Dvoretsky, and
  Ganichev]{Olbrich2013}
Olbrich,~P.; Zoth,~C.; Vierling,~P.; Dantscher,~K.-M.; Budkin,~G.~V.;
  Tarasenko,~S.~A.; Bel'kov,~V.~V.; Kozlov,~D.~A.; Kvon,~Z.~D.;
  Mikhailov,~N.~N.; Dvoretsky,~S.~A.; Ganichev,~S.~D. Giant photocurrents in a
  Dirac fermion system at cyclotron resonance. \emph{Phys. Rev. B}
  \textbf{2013}, \emph{87}, 235439\relax
\mciteBstWouldAddEndPuncttrue
\mciteSetBstMidEndSepPunct{\mcitedefaultmidpunct}
{\mcitedefaultendpunct}{\mcitedefaultseppunct}\relax
\EndOfBibitem
\bibitem[Bel’kov \latin{et~al.}(2005)Bel’kov, Ganichev, Ivchenko,
  Tarasenko, Weber, Giglberger, Olteanu, Tranitz, Danilov, Schneider,
  Wegscheider, Weiss, and Prettl]{Belkov2005}
Bel’kov,~V.~V.; Ganichev,~S.~D.; Ivchenko,~E.~L.; Tarasenko,~S.~A.;
  Weber,~W.; Giglberger,~S.; Olteanu,~M.; Tranitz,~H.~P.; Danilov,~S.~N.;
  Schneider,~P.; Wegscheider,~W.; Weiss,~D.; Prettl,~W. Magneto-gyrotropic
  photogalvanic effects in semiconductor quantum wells. \emph{J. Phys. Cond.
  Matt.} \textbf{2005}, \emph{17}, 3405\relax
\mciteBstWouldAddEndPuncttrue
\mciteSetBstMidEndSepPunct{\mcitedefaultmidpunct}
{\mcitedefaultendpunct}{\mcitedefaultseppunct}\relax
\EndOfBibitem
\bibitem[Weston \latin{et~al.}(2020)Weston, Zou, Enaldiev, Summerfield, Clark,
  Z{\'{o}}lyomi, Graham, Yelgel, Magorrian, Zhou, Zultak, Hopkinson, Barinov,
  Bointon, Kretinin, Wilson, Beton, Fal'ko, Haigh, and Gorbachev]{Weston2020}
Weston,~A. \latin{et~al.}  Atomic reconstruction in twisted bilayers of
  transition metal dichalcogenides. \emph{Nat. Nanotechnol.} \textbf{2020}.
  \relax
\mciteBstWouldAddEndPunctfalse
\mciteSetBstMidEndSepPunct{\mcitedefaultmidpunct}
{}{\mcitedefaultseppunct}\relax
\EndOfBibitem
\bibitem[Sturman and Fridkin(1992)Sturman, and Fridkin]{Sturman1992}
Sturman,~B.~I.; Fridkin,~V.~M. \emph{The Photovoltaic and Photorefractive
  Effects in Non-Centrosymmetric Materials}; Gordon and Breach Science
  Publishers: New York, 1992\relax
\mciteBstWouldAddEndPuncttrue
\mciteSetBstMidEndSepPunct{\mcitedefaultmidpunct}
{\mcitedefaultendpunct}{\mcitedefaultseppunct}\relax
\EndOfBibitem
\bibitem[Ivchenko(2005)]{Ivchenko2005}
Ivchenko,~E.~L. \emph{Optical Spectroscopy of Semiconductor Nanostructures};
  Alpha Sci. Int. Ltd.: Harrow, 2005\relax
\mciteBstWouldAddEndPuncttrue
\mciteSetBstMidEndSepPunct{\mcitedefaultmidpunct}
{\mcitedefaultendpunct}{\mcitedefaultseppunct}\relax
\EndOfBibitem
\bibitem[Belinicher and Sturman(1980)Belinicher, and Sturman]{Belinicher1980}
Belinicher,~V.~I.; Sturman,~B.~I. The photogalvanic effect in media lacking a
  center of symmetry. \emph{Sov. Phys. Usp} \textbf{1980}, \emph{23}, 199,
  [\textit{Usp. Fiz. Nauk} \textbf{1980}, \textit{130}, 415]\relax
\mciteBstWouldAddEndPuncttrue
\mciteSetBstMidEndSepPunct{\mcitedefaultmidpunct}
{\mcitedefaultendpunct}{\mcitedefaultseppunct}\relax
\EndOfBibitem
\bibitem[Olbrich \latin{et~al.}(2014)Olbrich, Golub, Herrmann, Danilov, Plank,
  Bel'kov, Mussler, Weyrich, Schneider, Kampmeier, Gr\"utzmacher, Plucinski,
  Eschbach, and Ganichev]{Olbrich2014}
Olbrich,~P.; Golub,~L.~E.; Herrmann,~T.; Danilov,~S.~N.; Plank,~H.;
  Bel'kov,~V.~V.; Mussler,~G.; Weyrich,~C.; Schneider,~C.~M.; Kampmeier,~J.;
  Gr\"utzmacher,~D.; Plucinski,~L.; Eschbach,~M.; Ganichev,~S.~D.
  Room-Temperature High-Frequency Transport of Dirac Fermions in Epitaxially
  Grown Sb$_2$Te$_3$- and Bi$_2$Te$_3$-Based Topological Insulators.
  \emph{Phys. Rev. Lett.} \textbf{2014}, \emph{113}, 096601\relax
\mciteBstWouldAddEndPuncttrue
\mciteSetBstMidEndSepPunct{\mcitedefaultmidpunct}
{\mcitedefaultendpunct}{\mcitedefaultseppunct}\relax
\EndOfBibitem
\bibitem[Weber \latin{et~al.}(2008)Weber, Golub, Danilov, Karch, Reitmaier,
  Wittmann, Bel'kov, Ivchenko, Kvon, Vinh, van~der Meer, Murdin, and
  Ganichev]{Weber2008}
Weber,~W.; Golub,~L.~E.; Danilov,~S.~N.; Karch,~J.; Reitmaier,~C.;
  Wittmann,~B.; Bel'kov,~V.~V.; Ivchenko,~E.~L.; Kvon,~Z.~D.; Vinh,~N.~Q.;
  van~der Meer,~A. F.~G.; Murdin,~B.; Ganichev,~S.~D. Quantum ratchet effects
  induced by terahertz radiation in GaN-based two-dimensional structures.
  \emph{Phys. Rev. B} \textbf{2008}, \emph{77}, 245304\relax
\mciteBstWouldAddEndPuncttrue
\mciteSetBstMidEndSepPunct{\mcitedefaultmidpunct}
{\mcitedefaultendpunct}{\mcitedefaultseppunct}\relax
\EndOfBibitem
\bibitem[Deyo \latin{et~al.}(2009)Deyo, Golub, Ivchenko, and Spivak]{Deyo2009}
Deyo,~E.; Golub,~L.~E.; Ivchenko,~E.~L.; Spivak,~B. Semiclassical theory of the
  photogalvanic effect in non-centrosymmetric systems. \emph{arxiv:0904.1917}
  \textbf{2009}. \relax
\mciteBstWouldAddEndPunctfalse
\mciteSetBstMidEndSepPunct{\mcitedefaultmidpunct}
{}{\mcitedefaultseppunct}\relax
\EndOfBibitem
\bibitem[Moore and Orenstein(2010)Moore, and Orenstein]{Moore2010}
Moore,~J.~E.; Orenstein,~J. Confinement-Induced Berry Phase and
  Helicity-Dependent Photocurrents. \emph{Phys. Rev. Lett.} \textbf{2010},
  \emph{105}\relax
\mciteBstWouldAddEndPuncttrue
\mciteSetBstMidEndSepPunct{\mcitedefaultmidpunct}
{\mcitedefaultendpunct}{\mcitedefaultseppunct}\relax
\EndOfBibitem
\bibitem[Lu \latin{et~al.}(2020)Lu, Lian, Chaudhary, Piot, Romagnoli, Watanabe,
  Taniguchi, Poggio, MacDonald, Bernevig, and Efetov]{Lu2020}
Lu,~X.; Lian,~B.; Chaudhary,~G.; Piot,~B.~A.; Romagnoli,~G.; Watanabe,~K.;
  Taniguchi,~T.; Poggio,~M.; MacDonald,~A.~H.; Bernevig,~B.~A.; Efetov,~D.~K.
  Fingerprints of fragile topology in the Hofstadter spectrum of twisted
  bilayer graphene close to the second magic angle. \emph{arxiv} \textbf{2020}.
  \relax
\mciteBstWouldAddEndPunctfalse
\mciteSetBstMidEndSepPunct{\mcitedefaultmidpunct}
{}{\mcitedefaultseppunct}\relax
\EndOfBibitem
\bibitem[Ziemann \latin{et~al.}(2000)Ziemann, Ganichev, Prettl, Yassievich, and
  Perel]{Ziemann2000}
Ziemann,~E.; Ganichev,~S.~D.; Prettl,~W.; Yassievich,~I.~N.; Perel,~V.~I.
  Characterization of deep impurities in semiconductors by terahertz tunneling
  ionization. \emph{J. Appl. Phys.} \textbf{2000}, \emph{87}, 3843--3849\relax
\mciteBstWouldAddEndPuncttrue
\mciteSetBstMidEndSepPunct{\mcitedefaultmidpunct}
{\mcitedefaultendpunct}{\mcitedefaultseppunct}\relax
\EndOfBibitem
\bibitem[Saleh and Teich(2007)Saleh, and Teich]{Saleh2007}
Saleh,~B.; Teich,~M. \emph{Fundamentals of photonics}; Wiley-Interscience:
  Hoboken, N.J, 2007\relax
\mciteBstWouldAddEndPuncttrue
\mciteSetBstMidEndSepPunct{\mcitedefaultmidpunct}
{\mcitedefaultendpunct}{\mcitedefaultseppunct}\relax
\EndOfBibitem
\end{mcitethebibliography}

\end{document}